# MOMENT QUADRIPOLAIRE
# MESURÉ PAR RÉSONANCE QUADRIPOLAIRE NUCLEAIRE
# PRINCIPE ET DÉFINITION

*Par* Mohamed **BELFKIR**

**Département de PHYSIQUE, Faculté des Sciences AGADIR**

**Université IBN ZOHR**

**MAROC**

(version 2)

*Résumé.*

*Le moment quadripolaire nucléaire est un caractère fondamental associé au noyau. Ce moment est lié à la répartition non purement sphérique au sein du noyau, en effet sa mesure nous permet de sonder la déformation géométrique du noyau de sa forme sphérique. Les méthodes de mesure du moment quadripolaire consiste à étudie l'énergie d'interaction hyperfine électrique entre le moment quadripolaire et le gradient du champ électrique dû aux électrons atomiques. L'une des méthodes est la résonance quadripolaire nucléaire RQN qui consiste à observe les transitions entre les niveaux d'énergies éclatées par l'effet de l'interaction quadripolaire et induites par un champ radiofréquence.*

*Abstract.*

*The nuclear quadruple moment is a fundamental character associated to the nuclei, this moment is related to the not purely spherical distribution in the nuclei, indeed its measure allows us to survey the geometric deformation of the nuclei of its spherical shape. The measurement methods of the quadruple moment is to study the electrical energy hyperfine interaction between the quadruple moment and the electric field gradient due to atomics electrons, one of the methods is the nuclear quadruple resonance NQR which is observed at the transitions between energy levels splits by the effect of the quadruple interaction and induced by a radio frequency field.*

**Introduction**

Les états nucléaires sont caractérisés par un moment électromagnétique nucléaire, ce moment électromagnétique se décompose en un moment magnétique et un moment multipolaire électrique.

À la distribution spatial des charges à l'intérieur d'un noyau, associé un vecteur multipolaire électrique composé de plusieurs termes monopolaire, dipolaire, quadripolaire et etc… Le terme monopolaire n'est rien que la charge totale du noyau, le moment dipolaire est nul pour tous les noyaux puisque le barycentre du noyau coïncide avec le centre des charges. On s'intéresse au moment quadripolaire, ce moment quadripolaire nucléaire est lié à la répartition des charges non purement sphérique au sien du noyau, sa mesure nous permet de sonder la déformation géométrique du noyau selon un axe d'observation, elle délivre ainsi des informations sur la structure collective du noyau dans l'état mesuré.

Le but de cette mise au point est d'étudie le moment quadripolaire nucléaire et sa mesure par résonance quadripolaire nucléaire RQN. Nous traiterons dans ce manuscrit le principe de la théorie de RQN après une définition du moment quadripolaire nucléaire relativement simple qu'est traduit le concept de la déformation géométrique de la forme sphérique du noyau.

### I. Moment quadripolaire nucléaire

La distribution de charges à l'intérieur d'un noyau, peut être étudie par l'intermédiaire de son interaction avec le champ électrique crée par le cortège électronique du noyau. L'énergie de cette interaction peut se mettre sous la forme suivante :

$$E = \iiint V \rho dv \quad (1)$$

Dans cette expression $V$ représente le potentiel dériver par le champ électrostatique qui règne dans l'espace au se trouver le noyau, à l'origine égale à $V_0$, $\rho$ est la densité des charges nucléaires à l'intérieur du noyau, $dv$ est un élément de volume du noyau.

Toutes les méthodes du calcul du moment quadripolaire nucléaire consistent à développer l'énergie d'interaction $E$. Tant que le volume du noyau est très petit on développe cette énergie en série de Taylor au voisinage de zéro et on se limite à l'ordre deux, le développement donne [1] :

$$E = V_0 \iiint dq + \frac{\partial V}{\partial x}\bigg|_0 \iiint x dq + \frac{\partial V}{\partial y}\bigg|_0 \iiint y dq + \frac{\partial V}{\partial z}\bigg|_0 \iiint z dq +$$
$$\left[\frac{1}{2}\left\{\left(\frac{\partial^2 V}{\partial x^2}\right)_0 \iiint x^2 dq + \left(\frac{\partial^2 V}{\partial y^2}\right)_0 \iiint y^2 dq + \left(\frac{\partial^2 V}{\partial z^2}\right)_0 \iiint z^2 dq\right\} + \left(\frac{\partial^2 V}{\partial x \partial y}\right)_0 \iiint xy dq + \left(\frac{\partial^2 V}{\partial y \partial z}\right)_0 \iiint yz dq + \left(\frac{\partial^2 V}{\partial x \partial z}\right)_0 \iiint xz dq\right] + \ldots \quad (2)$$

Le terme entre crochets est le terme quadripolaire qui traduit l'énergie d'interaction due au moment quadripolaire.

On impose au noyau d'être un ellipsoïde de révolution, et on admettre que l'axe de révolution est l'axe qui porte le vecteur du spin nucléaire $\vec{I}$, par un calcul relativement simple on simplifie ce terme qui peut s'écrit :

$$E_Q = \frac{1}{4}\left(\frac{\partial^2 V}{\partial z^2}\right)_0 \iiint (3z^2 - r^2)\rho dv \quad (3)$$

Avec $r^2 = x^2 + y^2 + z^2$

À ce niveau on définit le moment quadripolaire intrinsèque $Q_I$ par :

$$Q_I = \frac{1}{e}\iiint (3z^2 - r^2)\rho dv \quad (4)$$

Cette définition permet de donner au moment quadripolaire la dimension de carré de longueur, étant donné son ordre de grandeur nous l'exprimerons en « barn ».

Le moment quadripolaire intrinsèque ne correspond plus au moment quadripolaire observé $Q$, puisque le vecteur du spin nucléaire n'est pas généralement porté par l'axe d'observation.

Pour une valeur observable du moment quadripolaire $Q$ positive, le noyau est un ellipsoïde allongé suivant l'axe d'observation, pour une valeur négative l'ellipsoïde est aplati vis-à-vis de ce même axe et lorsque la valeur observée est nulle la répartition des charges est purement sphérique (figure suivante [5]).

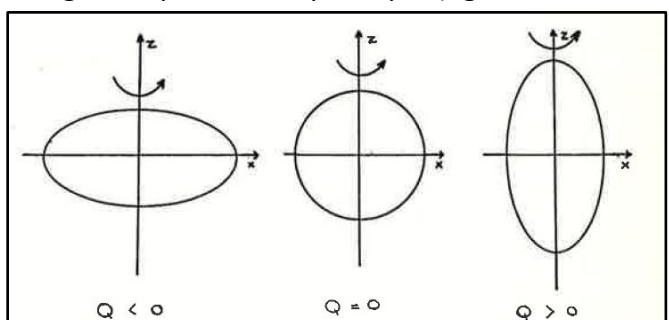

**Figure 1 : déformation du noyau selon la valeur du moment quadripolaire**

## II. Interaction hyperfine électrique

Lors de la présence d'un noyau quadripolaire (noyau du spin supérieur strictement à ½) dans un système matériel, le gradient du champ électrique dû aux électrons atomiques est susceptible de se coupler avec le moment quadripolaire de ce noyau, l'interaction de couplage (couplage quadripolaire) ou l'interaction hyperfine électrique permette de séparer les niveaux d'énergie quadripolaire de système considéré.

### 1. Gradient de champ électrique

Le moment quadripolaire ne peut se coupler qu'avec un champ électrique inhomogène tant que l'action du champ électrique homogène sur le moment quadripolaire est nulle [4], ce gradient du champ se représente dans système d'axes cartésiens arbitrairement choisis par un tenseur symétrique de rang 2 :

$$\begin{pmatrix} V_{xx} & V_{xy} & V_{xz} \\ V_{yx} & V_{yy} & V_{yz} \\ V_{zx} & V_{zy} & V_{zz} \end{pmatrix} \quad (5)$$

$V_{xx} = \frac{\partial E}{\partial x} = \frac{\partial^2 V}{\partial x^2}$

On peut toujours trouver un référentiel moléculaire dont les axes de système sont les axes principaux du tenseur de gradient du champ électrique, donc [2],[3]:

$$\begin{pmatrix} V_{xx} & 0 & 0 \\ 0 & V_{yy} & 0 \\ 0 & 0 & V_{zz} \end{pmatrix} \quad (6)$$

Les axes de référentiel sont choisis tel sort que :

$$|V_{zz}| \geq |V_{yy}| \geq |V_{xx}| \quad (7)$$

L'équation de Laplace nous permet de déduire que la trace du tenseur est nulle, le fait que la trace est nulle a pour conséquence que le couplage quadripolaire ne conduit à aucune éclatement des niveaux d'énergies en phase liquide tant que la valeur mesurable de gradient de champ le long d'une interaction anisotrope est le tiers de la trace du fait des réorientations moléculaires rapides.

L'invariante de la trace nous permet d'introduire pour un système matériel solide le paramètre d'asymétrie définit par :

$$\eta = \frac{V_{xx} - V_{yy}}{V_{zz}} \quad (8)$$

Le paramètre d'asymétrie est intimement lié à la symétrie de la maille cristalline dans laquelle se cristallise le noyau quadripolaire étudie, donc la symétrie du champ, les mailles hexagonales ont une symétrie axiale donc un paramètre d'asymétrie nul, alors que les mailles trigonales sont généralement asymétriques, les mailles cubiques octaédriques ne représentent aucun gradient de champ électrique.

La constante du couplage $V_{zz}$ et le paramètre d'asymétrie sont spécifiques de l'environnement électronique du noyau considéré, ils représentent une empreinte digitale très précise de la molécule analysée, et suffisants pour bien définit le gradient de champ.

## 2. Hamiltonien d'interaction hyperfine électrique

Le système étudie est microscopique donc l'étude se fait au moyen de la mécanique quantique, l'énergie du couplage quadripolaire va décrit par l'Hamiltonien d'interaction hyperfine électrique [2],[3] :

$$\mathcal{H}_Q = \frac{eQ}{6I(2I-1)}\left[(3I^2_x - I^2)V_{xx} + (3I^2_y - I^2)V_{yy} + (3I^2_z - I^2)V_{zz}\right]$$

Avec $\vec{I}$ l'opérateur vectoriel du spin nucléaire
Ou encore on utilise les paramètres $V_{zz}$ et $\eta$

$$\mathcal{H}_Q = \frac{eQV_{zz}}{4I(2I-1)}\left[(3I^2_z - I^2) + \eta(I^2_x - I^2_y)\right] \quad (9)$$

On peut toujours arriver au même résultat on utilisant une expression plus simple d'Hamiltonien quadripolaire [5] :

$$\mathcal{H}_Q = \frac{eQ}{2I(2I-1)}\vec{I}V\vec{I} \quad (10)$$

On est en droit de se demander pourquoi le couplage quadripolaire peut être appréhendé comme un couplage magnétique de spin alors qu'il est résulté d'une interaction purement électrique. On évite les développements complexes et fastidieux, On peut s'attaché à une approche plus simple que les raisonnements que l'on peut trouver sur cette question, nous pouvant assimiler, rappelons-le, le moment quadripolaire à une distribution des protons au sein du noyau. Or cette distribution est éventuellement liée à celle des spins. Il semble donc normale que les opérateurs de spins interviennent dans l'expression de l'énergie d'interaction.

## III. Résonance Quadripolaire Nucléaire RQN

Les méthodes de mesure de moment quadripolaire consiste à étudie l'énergie d'interaction quadripolaire, l'une des méthodes est la résonance quadripolaire nucléaire, cette méthode spectroscopique est similaire en principe à la résonance magnétique nucléaire RMN qui très célèbre chez les chimistes, RQN consiste à observer les transitions entre les niveaux d'énergies éclatées par l'interaction hyperfine électrique et induits par un champ radiofréquence $\vec{B}$.

En principe les deux méthodes spectroscopiques sont généralement comparables sauf des différences au niveau de sensibilité et domaine d'application ainsi que la popularité. Comme la RQN est faiblement sensible et n'est pas très répandu et très difficile à détecter nous impose une difficulté de son utilisation effective, contrairement à la RMN qu'est détectable. les deux méthode repose sur l'observation des transitions entre les niveaux d'énergies éclatées induit par un champ radiofréquence, un champ magnétique statique est nécessaire dans la RMN c'est l'origine de l'éclatement des niveaux d'énergies par effet Zeeman, c'est encore une différence qui vient à l'origine de l'éclatement des niveaux d'énergies puisque la RQN a pour origine l'interaction quadripolaire entre le gradient de champ électrique et le moment quadripolaire induit par la distribution non sphérique des charges électrique au sein du noyau. RQN et RMN permettent de détecté les noyaux des spins non nul et comme la RMN est en théorie capable de détecté tous les noyaux possédant un spin nucléaire non nul dans les phases solide et liquide, malheureusement elle ne permette pas d'analyse les noyaux quadripolaires qui représente le ¾ des éléments de tableau périodique en phase solide.

## IV. Application : détection de l'azote 14 dans le perchlorate d'ammonium $NH_4ClO_4$

Dans cette partie on s'intéresse d'appliqué tous les résultats obtenu jusqu'à présent dans le cas de l'azote 14 qu'est un noyau quadripolaire de spin nucléaire égale à l'unité, l'Hamiltonien quadripolaire s'écrit dans ce cas :

$$\mathcal{H}_Q = \frac{w_Q}{3}\left[3I^2_z - 2 + \eta(I^2_x - I^2_y)\right] \quad (11)$$

Avec $w_Q = \frac{3eQV}{4}$ est la fréquence quadripolaire.

La matrice représentative de $\mathcal{H}_Q$ dans la base des états de l'opérateur $I_z$ est :

$$\mathcal{H}_Q = \frac{w_Q}{3}\begin{pmatrix} 1 & 0 & \eta \\ 0 & -2 & 0 \\ \eta & 0 & 1 \end{pmatrix} \quad (12)$$

La détermination des niveaux d'énergie induite par l'interaction du couplage consiste la détermination des valeurs et vecteurs propres de l'Hamiltonien d'interaction dans les deux cas d'une symétrie purement axiale et une symétrie quelconque ($\eta \neq 0$), par un calculer relativement simple de la mécanique quantique on arrive aux résultats que l'on regroupe dans le diagramme énergétique suivant [2],[3] :

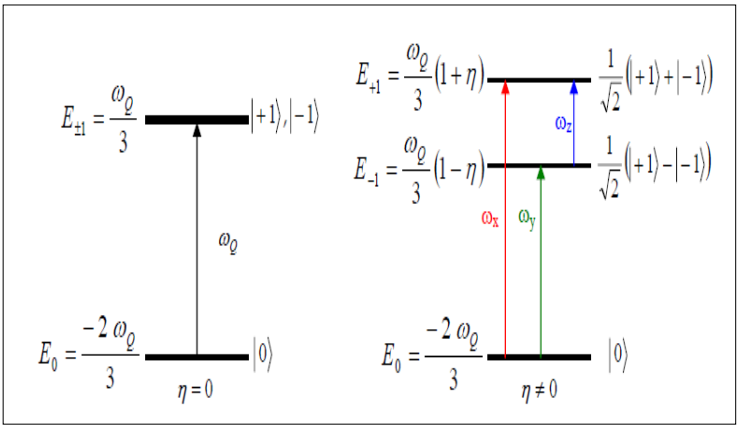

Figure 2 : Diagramme énergétique (I=1)

Dans le cas d'une symétrie axiale le couplage quadripolaire induit deux niveaux d'énergies l'un est doublement dégénéré, contrairement au cas au la maille cristalline ne représente aucun symétrie bien définit les niveaux d'énergie induites par le couplage dans ce cas sont trois niveaux simples.

On s'intéresse maintenant à mesuré ce couplage par RQN donc à mesuré le moment quadripolaire de l'azote 14, on perturbe notre système par un champ radiofréquence et on déduit les fréquences des transitions entre les niveaux d'énergies, pour le cas d'une symétrie axiale une seule transition possible exactement à la fréquence du couplage quadripolaire dans le cas d'une symétrie non axiale les fréquences des transitions sont :

$w_x = w_Q\left(1 + \frac{\eta}{3}\right), w_y = w_Q\left(1 - \frac{\eta}{3}\right), w_z = \frac{2w_Q}{3}\eta$

Une mesure de moment quadripolaire de l'azote 14 dans le perchlorate d'ammonium donne comme résultat le spectre de raies suivant [2]:

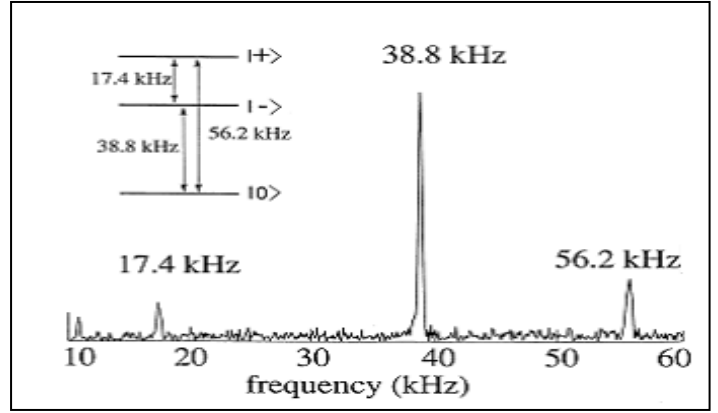

Figure 3: Spectre de RQN de l'azote 14 dans le perchlorate d'ammonium

D'après le spectre :
$w_x = 56.2\ kHz$, $w_y = 38.8\ kHz$ et $w_z = 17.4\ kHz$

À la fin de cet article, on peut dire que tous les astuces de la définition et calcul du moment quadripolaire nucléaire sont baser sur le développement de l'énergie d'interaction de la charge du noyau avec le potentiel dû aux électrons atomiques, et peut se mesure à l'aide de plusieurs méthodes tels que l'orientation nucléaire a base température et le mélange des niveaux nucléaires [6]… etc.

## Bibliographie

[1] D.BLANC. *physique nucléaire.* MASSON

[2] M.FERRARI. thèse. *aspects fondamentaux de la résonance quadripolaire nucléaire de l'azote 14 par impulsion de champ radiofréquence vérification expérimentales.* Nancy Université.

[3] S.AISSANI. thèse. *La résonance quadripolaire nucléaire de l'azote 14 amélioration de la qualité spectrale et effet d'un champ magnétique statique de faible amplitude.* Université de Loraine.

[4] J.COUSSEUA. *la spectroscopie de résonance quadripolaire nucléaire Principe et application en chimie organique.* Partie 1. Labo de chimie organique. Institut de recherches scientifiques et techniques. Université d'Argens

[5] S.ASHBROOK. *introduction to quadruple NMR.* School of chemistry. University of St Andrews

[6] B.HLIMI. thèse. *Moments quadripolaires déterminés par orientation nucléaire.* Université Coulde Bernard Lyon 1